\documentstyle[12pt]{report}
\begin{document}
\pagestyle{empty}

\begin{center}
{\Large Charge Transfer Excitons and the Metal-Insulator Transition in the
High Temperature Superconductors.}
\end{center}
\vspace{0.4in}
\begin{center}
C. Vermeulen$^{1}$, W. Barford$^{1}$ and E. R. Gagliano$^{2*}$
\end{center}

\begin{center}
1. Department of Physics, The University of
Sheffield, Sheffield, S3 7HR, United Kingdom.\\

2. Physics Department, University of Illinois at Urbana-Champaign,
1110 W. Green, Urbana, Il 61801, U.S.A. \\

\end{center}
\vspace{1.0in}
\begin{center}
Abstract
\end{center}
The effects of the nearest neighbour Coulomb repulsion, $V$, are
 considered in the one dimensional copper-oxide chain using the modified
Lanczos
 method.  Above a critical value of $V$ we find that charge transfer excitons
  are the lowest lying energy exitations in the insulating phase.
Close to the metal-insulator transition and $V=0$ there is good `particle-hole'
symmetry, showing that the mapping from a two band model to a one band model is
appropriate.
The effect of the nearest neighbour repulsion is to create states in the
gap and to
pin the Fermi level upon hole doping.
\\
\\
$^*$Present address: Centro Atomico, Bariloche, 8400 Bariloche, Argentina.

\pagebreak

\topmargin 0.0in
\footheight 0.5in
\footskip 0.5in
\oddsidemargin 0.0in
\textwidth 6.0in
\textheight 9.0in

There are many features of the high temperature superconductors which cannot be
explained within a one band Hubbard model. Two examples of these are: first,
the excitonic excitations observed recently by Raman spectroscopy \cite{lue93}
and second, the pinning of the Fermi level
as a function of doping, which has long been an outstanding problem presented
by photoemission spectroscopy \cite{Allen90,Anderson93}.
It is the purpose of this letter to show that both of these phenomena occur
as a natural consequence of the Coulomb repulsion between copper and oxygen
orbitals.
 To demonstrate this we solve exactly finite size chains via the Lanczos
method. We also provide physical arguments to support these calculations.

Our model for the copper-oxide chain consists of two atoms per unit
cell.  Neglecting the oxygen-oxygen hybridization and considering the Coulomb
interaction up to first-nearest neighbours, the copper-oxide chain is
described by the two band model Hamiltonian,
$$
H = H_t + {\Delta \over 2} \sum_{ij\sigma}{(p^{\dagger}_{j\sigma}p_{j\sigma}
    -d^{\dagger}_{i\sigma}d_{i\sigma})}
        +  U_{d} \sum_{i} d^{\dagger}_{i\uparrow}d_{i\uparrow}
        d^{\dagger}_{i\downarrow}d_{i\downarrow}
        +  U_{p} \sum_{j} p^{\dagger}_{j\uparrow}p_{j\uparrow}
        p^{\dagger}_{j\downarrow}p_{j\downarrow}\nonumber\\
$$
$$
        +  V \sum_{<ij>\sigma\sigma^\prime}
\left(d^{\dagger}_{i\sigma}d_{i\sigma}
             p^{\dagger}_{j\sigma^\prime}p_{j\sigma^\prime}\right)
         +\left( {(U_d+U_p) \over 4} + 2V \right)
         (N_{s}-N_{p}),
$$
where
$$
\hspace{3cm} H_t = -t \sum_{<ij>\sigma} {(d^{\dagger}_{i\sigma}
p_{j\sigma} + h.c.). }
\hspace{3cm} (1)
$$
\noindent $i$ and $j$ are copper and oxygen sites
respectively,  $<ij>$ represents nearest neighbours and the operator
$d^{\dagger}_{i\sigma}~~(p^{\dagger}_{j\sigma})$ creates a $Cu$ $(O)$ hole with
spin $\sigma$.  $\Delta$ is the charge-transfer energy, $U_d$
$(U_p)$ is the copper (oxygen) Coulomb repulsion, $N_s$ is the number of sites,
$N_{p}$ is the number of particles, and $V$ and $t$ are the
copper-oxide Coulomb repulsion and hybridization, respectively.  The
Hamiltonian
has been defined so that it is invariant under the `particle-hole
transformation' $d^{\dagger}_{i\sigma} \to d_{i\sigma}$, $p^{\dagger}_{j\sigma}
\to p_{j\sigma}$ and $\Delta \to U_d-(\Delta+U_p)$, {\it i.e.} E($N_{p},
\Delta$)= E($2N_{s}-N_{p},U_d-(\Delta+U_p)$).

In the non-interacting limit the Hamiltonian
can be solved exactly.  Two bands are formed with a charge transfer gap between
them, $\epsilon_k = \pm\left[ {\Delta^{2} \over 4} + {4 t^2 } \cos^2 (k/2)
\right]^{1 \over 2}.$  The diagonalised eigenfunctions corresponding to the
lower and upper bands are $\alpha_{\sigma}^{1\dagger}= u_1
d_{i\sigma}^\dagger + v_1 p_{j\sigma}^\dagger$ and
$\alpha{_\sigma}^{2\dagger}= u_1 d_{i\sigma}^\dagger - v_1
p_{j\sigma}^\dagger$, respectively where $u_1 = {\sqrt{{|\epsilon|+ \Delta/2}
\over 2|\epsilon|}}$ and $v_1 = {\sqrt{{|\epsilon|- \Delta/2} \over
2|\epsilon|}}$.
In the strong coupling limit, $U_{p},U_{d}\rightarrow
\infty$, the spin label becomes redundant and the Hamiltonian maps onto a
spinless Fermion model \cite{emery90}.  When $V=0$, the eigenfunctions are as
before, except that there is no spin index.  Consequently, $k_f \to 2k_f$ and
the system is Mott insulating for one hole per unit cell.

When the infinite $U$ limit is relaxed the spin degeneracy is lifted by the
superexchange interaction. There is spin and charge separation with the spin
excitations described by a `squeezed' Heisenberg chain and the charge
excitations
by a Slater determinant of spinless fermions.

In this letter we consider parameters which are more relevant to the
high temperature superconductors. Thus we take $U_d\sim8t$, $\Delta\sim2-3t$
and $U_{p}$
small, where $t\sim1.5eV$. However, as we are interested in the role played by
the nearest neighbour interaction, $V$, this will be left as a free parameter.
\\

Firstly we examine the model in the insulating phase at a hole density of
$n=0.5$.  At this density  the system is a charge transfer insulator for all
$\Delta\neq0$, and the ground state is a spin density wave.
 The majority of the charge resides on the copper sites, forming a
N\'eel state with a momentum of $k=\pi$ ({\it a wavelength of two unit cells}).

As $V$ is increased there is a change in the
character of the low lying energy excitations in the charge transfer gap, being
particle-hole for small $V$ and excitonic like above $V \sim 1.5t$.  The
exciton
is identified as the lowest frequency peak, $E_{e}$, of the dynamical
current-current correlation function, $J(w)$, when it appears inside the
charge transfer gap,
 $E_g$.  $J(w)$ is defined by
\begin{eqnarray}
J(w)=-\frac{1}{\pi}Im(G^{R}(w)), \nonumber
\end{eqnarray}
where
\begin{eqnarray} \hspace{1cm}
G^{R}(w) &=& F.T.<\psi_{0}\mid j^{\dagger}(t)j(0)\mid\psi_{0}>  \nonumber\\
         &=& <\psi_{0}\mid j^{\dagger}(0)\left(
\frac{1}{H-(E_{0}+w)+i\eta}\right)j(0)\mid\psi_{0}>_{\eta\rightarrow0}.
\nonumber \hspace{1cm}(2)
\end{eqnarray}
$G^{R}(w)$ is the retarded Green function and $j$ is the current operator
defined as
     $j(0)=-i\sum_{l\sigma}(c^{\dagger}_{l\sigma}
     c_{l+1\sigma}-c^{\dagger}_{l+1\sigma}c_{l\sigma})$. $\{\mid\psi_{0}>,
E_0\}$ are the ground state
     eigensolutions. In practice
       we choose a finite value of $\eta$ to broaden the peaks.  The Green
function
     is calculated via the continued fraction \cite{Gag89},
\begin{eqnarray} \hspace{3cm}
G^{R}(w)
     =\frac{<\psi_{0}\mid j^{\dagger}j\mid\psi_{0}>}{w-a_{0}
     -\frac{b^{2}_{1}}{w-a_{1}-\frac{b^{2}_{2}}{w-a_{2} -\ldots}}}. \nonumber
\hspace{5cm} (3)
\end{eqnarray}

 To find the coefficients $\left\{a_{n},b_{n}^{2}\right\}$ we use a Lanczos
type algorithm as follows.
\begin{quotation}
(1) Define the state $\mid f_{0}>=j\mid \psi_{0}>$.

 (2) Generate a set of orthonormal states by the relationship,
 \end{quotation}
 \begin{eqnarray}
   \mid f_{n+1}> &=& H\mid f_{n}> -a_{n}\mid f_{n}> - b^{2}_{n}\mid f_{n-1}>,
 \nonumber
 \end{eqnarray}
 \begin{quotation}
   where
 \end{quotation}
 \begin{eqnarray} \hspace{3cm}
 a_{n} &=& <f_{n}\mid H\mid f_{n}>/<f_{n}\mid f_{n}>, \nonumber \\
 b^{2}_{n+1} &=& <f_{n+1}\mid f_{n+1}>/<f_{n}\mid f_{n}> \nonumber\\
\mbox{and}\hspace{3cm}   b^{2}_{0} &=& 0. \nonumber \hspace{6.5cm} (4)
\end{eqnarray}

Figure 1 is a
 typical profile of $J(w)$.  The inset shows how the energies
 $E_e$ and $E_g$ evolve
as $V$ is increased, and also the cross-over point which occurs
 at around $V\sim1.5t$.  The exciton may be pictured as a hole that has
 been excited from the lower Hubbard band (or valence band), which is
predominately copper in character, to the conduction band, which is
predominately oxygen in character.
  The real space picture is therefore of a hole on a
copper site hopping onto a neighbouring oxygen site. This effectively leaves an
`electron' on the copper site and a hole on the oxygen site and due to the
Coulomb repulsion between neighbouring sites this leads to an effective
attraction between the electron and the hole. This attraction reduces the
energy to a point below the charge transfer gap and causes the formation of a
tightly bound `electron-hole' pair, also known as a Frenkel exciton.
The energy in forming the exciton is a balance between the Coulomb attraction
of the `electron-hole' pair and the kinetic energy loss.  The former is driven
by $V$, whereas the latter is determined by $t$.  Hence, when $V$ is of the
order of $t$ we expect excitons to exist within the charge transfer gap.
The existence of excitons in this model is an encouraging sign that the nearest
neighbour Coulomb repulsion may play an important role in the insulating
cuprates.  Recent Raman scattering experiments by Liu {\it et al}. \cite{lue93}
indicate the existence of excitons in these materials.  Their experiments were
carried out on a variety of cuprates in the insulating state.  They showed that
these materials have a complex collection of excitonic states.
Two possible explanations they give for this are either, an intra-atomic
transition $d_{x^{2}-y^{2}}\rightarrow d_{xy}$, or an inter-atomic transition
of
the kind $d_{x^{2}-y^{2}} \rightarrow p_{xy}$.  We appear to observe the
latter.
It would be interesting to know what role, if any, these charge transfer
excitons may have within the conducting region, and whether they could be a
possible mechanism for superconductivity.
\\

We now consider the effect of the Coulomb repulsion $V$ on the photoemission
spectra.  For finite $U_d$, but keeping $U_p,V=0$,
studies of the transfer of spectral weight reveal that at low doping there is
a good symmetry between the strongly correlated copper sites and the
`free' electron sites of oxygen.  The transfer of spectral weight is
calculated from the single particle Green function.  This is defined
in the same way as
equation (2), but with $j^{\dagger}$ replaced by $c^{\dagger}_{k\sigma}$;
the spectral function is then
given by $S_{\sigma}(w)=-\frac{1}{\pi}\sum_{k}Im(G_k^{R}(w))$.  Again, it is
evaluated using the modified Lanczos procedure.  Figure 2 shows the spectral
function for $6$ holes on a $12$ site chain with $V=0$.  The lower
Hubbard band, upper Hubbard band and the `oxygen' conduction band
are clearly evident.  The values
of the parameters are choosen to fit, as far as possible, with photoemission
data, which suggests $U_d\sim 8t-9t, \Delta\sim 2t-3t$ and $U_p$ small
\cite{Tun91}.

The transfer of spectral weight is a measure of the strength of the
correlations
as the hole density is varied. It is defined by
$S_{T}=\int^{e_{gap}}_{e_{f}} S_{\sigma}(w)dw$, where
$e_{gap}$ is the energy of the {\it nearest} gap to the Fermi
surface.  Figure 3 shows $S_{T}$ for a $12$ site chain with $V=0$.
For highly correlated systems the transfer of spectral weight goes as
roughly twice the doping fraction, as weight is transfered from both the upper
and the lower Hubbard bands to form new states at the Fermi level \cite{esk91}.
  It might
seem suprising, therefore, to find that doping into the conduction band
with holes
is symmetric with respect to doping into the lower Hubbard band with electrons.
However, the cuprates cannot be described as  normal semiconductors
because of the strong hybridization between copper and oxygen sites.  Near to
the metal-insulator transition the Zhang-Rice picture of an oxygen spin bound
to a copper spin as a singlet is appropriate \cite{Zhang88}.  This picture
naturally leads to a
    mapping from a two band model to a one band model, which is exact in the
limit $U_d-\Delta << U_d$ and small doping \cite{Barford90}.  Further from the
metal-insulator transition, however, correlation effects are screened by the
 added oxygen holes, the Zhang-Rice singlet breaks down and the transfer of
spectral weight becomes similar to that expected of a doped semiconductor.
This
effect can be interpreted in figure 3, where one can see a reduction in the
gradient of the line as the hole density approaches half filling ($n=1$).

A large value of the
nearest neighbour Coulomb interaction has the effect of destroying the charge
transfer gap on doping, so a comparison with the transfer of spectral weight is
impossible. However, for small $V$ there is very little change from the curve
in
figure 3, so it has not been shown.  Turning on the nearest neighbour
interaction has one very important effect,  however.  Experimental work on
samples of hole doped $La_{2-x}Sr_{x}CuO_{4-y}$
\cite{Allen90,Anderson93} has shown that there is a significant increase in
spectral weight within the charge transfer gap on hole doping. The Fermi
energy, $e_{f}$, has essentially the same position in the gap for both the
stochiometric and doped samples.
    These states must arise by a transfer of spectral weight from both the
conduction and valence band states of the insulator.  A similar effect is
observed in our calculations on the extended Hubbard model.  For $V=0$ hole
doping shifts the Fermi energy from the middle of the charge transfer gap into
the conducting band.  For $V\neq 0$, however, the effect of doping with
holes is to form new states in the charge transfer gap, resulting in the
pinning
of $e_{f}$.  Figure 4 shows the spectral weight for 8 holes on the
12 site chain with $U_d=8t$, $U_p=0$, $\Delta=2t$ and $V=2.5t$. The new
states which have
formed within the charge transfer gap are clearly evident.
Figure 5 is a plot
of $2e_{f}/E_g$ versus density $n$ for the values of $V=0$, $t$ and $2.5t$
for $0.5<n<1.0$, from which it is clear that the increase in $e_f$ on hole
doping decreases as $V$ is increased, and at $V=2.5t$ there is actually a
reduction in $e_f$ between $n=\frac{10}{12}$ and $n=1$.

The creation of new states in the gap and the pinning of $e_f$
can be thought of in the following manner. When $V=0$ the doping of holes onto
oxygen sites has no effect on the charge transfer gap, nor on the occupancy of
the copper sites.  When $V>0$, however, the doping of holes onto oxygen
sites has the effect of
pushing holes off copper sites onto vacant oxygen sites. This has two
consequences.
First, the nearest neighbour repulsion experienced by an oxygen hole decreases
because the amount of copper charge has been reduced.  Hence the effective
charge transfer gap shrinks. Second, new particle removal
states are formed which are oxygen in character. These result from the
copper holes
which have been pushed onto oxygen sites. The consequence of this is that new
particle addition states are formed which are copper in character.  This
admixture of the hole addition and removal states is responsible for the new
states in the gap. The pinning of the Fermi energy due to the nearest
neighbour repulsion also indicates a tendency towards phase separation
\cite{Barford93,Vermeulen94}. The fact that
pinning of $e_f$ is observed is further encouraging evidence that
the nearest neighbour repulsion may play an important role in cuprate
electron structure.
\\

 In conclusion,
we have shown that the nearest neighbour repulsion leads to
low lying excitons in the charge transfer gap at values of $V>1.5t$.
There is evidence from  recent Raman spectroscopy that these excitons are
indeed present in cuprate insulators. At $V=0$ there is good symmetry between
doping into the oxygen conduction band with holes, and doping into the copper
valence band with electrons. This indicates there are strong correlations
within the oxygen band and that the Zhang-Rice singlet is energetically
favourable. At finite values of the nearest neighbour repulsion
 doping with holes causes new states  to form in the charge transfer gap and
 results in the pinning of the
Fermi energy, as observed by photoemission spectroscopy.

\begin{center}
Acknowledgements.\\
\end{center}
We thank the SERC (United Kingdom) for the provision of a Visiting Research
Fellowship (ref. GR/H33091). W.B. also acknowledges support from the
SERC (ref. GR/F75445) and a grant from the University of Sheffield research
fund,
while E.G. acknowledges support from DMR89-20538/24. C.J.V. is supported by a
University of Sheffield scholarship.  We thank B. Alascio, A. Aligia,
S.  Bacci, C. Balseiro, E. Fradkin, G. A. Gehring and R. M. Martin for useful
conversations.  The computer calculations were performed at the National Center
 for Supercomputing Applications, Urbana, Illinois and at
 the University of Sheffield.

\pagebreak

\begin{figure}
{\bf Figure 1:} The current-current response function at $U_{d}=8t$,
$U_{p}=0$, $\Delta=2t$ and $V=2.5t$
for the stoichiometric chain. $E_g$ is the charge
transfer gap energy and $E_e$ identifies the excitonic low energy excitation.
The inset shows  $E_e$ and $E_g$ as a function of $V$.
\\
\\
{\bf Figure 2:} The single particle spectral function for 6 holes on a 12 site
 chain. This shows the lower and upper Hubbard bands and the
oxygen conduction band. $U_{d}=8t$, $U_{p}=0$, $\Delta=2t$ and $V=0$.
\\
\\
{\bf Figure 3:} The transfer of spectral weight as a function of doping for
$U_d=8t$,
  $U_p=0$, $\Delta=2t$ and $V=0$ on a 12 site chain.
 There is good symmetry between the highly correlated
copper band and the `electron-free' oxygen band. Notice that because of the
`particle-hole' symmetry of equation (1) $0.5\leq n \leq 1.0$ corresponds to
$Cu^{2+}\rightarrow Cu^{+}$ excitations, while $1.0\leq n \leq 1.5$
corresponds to $Cu^{2+}\rightarrow Cu^{3+}$ excitations.
\\
\\
{\bf Figure 4:} The single particle spectral function for 8 holes on a 12 site
 chains for $V=2.5t$. The effect of $V$ is to pin the Fermi surface near the
middle of the charge transfer gap and to create new states in the gap.
($U_{d}=8t$, $U_{p}=0$ and $\Delta=2t$).
\\
\\
{\bf Figure 5:} The plot of $2e_f/E_g$ versus density for three different
values of
$V$. The increase in $V$ has the effect of surpressing the
shift in $e_f$ on hole doping.
\end{figure}

\end{document}